What Can Learn from PER: Physics Education Research?

Chandralekha Singh, Department of Physics and Astronomy and Discipline-based Science Education Research Center (dB-SERC), University of Pittsburgh, Pittsburgh, PA 15260

Abstract (Note by Diane Riendeau, Column Editor for "For the New Teacher" of The Physics Teacher Journal): I believe that most teachers develop a belief in a set of pedagogical practices. As we teach, we try different ways to teach topics and then judge how successful the methods were. After several years, we have a compilation of techniques in our teaching toolbox. New teachers are at a disadvantage because they have fewer prior experiences to draw upon. Luckily, there is a group of physicists and physics educators who are researching how students learn physics, and have been able to show evidence of effective education practices in physics. They field of study is called PER: Physics Education Research. I asked Chandralekha Singh, one of the leaders in PER, to summarize some of the most relevant PER findings and her response follows.

**Overview:** Physics Education Research (PER) focuses on understanding how students learn physics at all levels and developing strategies to help students with diverse prior preparations learn physics more effectively. New physics instructors are encouraged to visit http://PhysPort.org, a website devoted to helping instructors find effective teaching resources based upon PER.[1] The site has links to PER-based teaching approaches, hosts instructional materials sorted by pedagogy, and provides assessment tools and tips for troubleshooting if hurdles are encountered in the implementation of PER-based instructional strategies.

**Interactive Engagement:** ~~All~~ Many PER-based curricula and pedagogies build on the key research finding that teaching by telling is not an effective approach to helping students learn physics[2]. Students must construct their own understanding, though the instructor plays a critical role in helping students accomplish this. An instructor should model the criteria of good performance, while leaving sufficient time to provide guidance and feedback to students as they practice useful skills. The amount of instructional support given to students should be decreased gradually as they develop self-reliance. One approach that has been shown to be effective if utilized with care is to let students work in small groups and take advantage of each other's strengths. Effective group work has been demonstrated to result in co-construction of knowledge[3]—that is, students working together can figure out answers to questions even without instructor's help that they would not have figured out on their own.

**Importance of building on students' initial knowledge:** Another important research finding is that students are not "blank slates"[4]. Thus, instructors should make an effort to learn the initial knowledge of their students at the beginning of the course in order to provide appropriate guidance and support to help them learn physics. For example, in an introductory physics course, students' naïve conceptions based upon their everyday experiences—motion implies net force, or in a collision, a truck must exert a larger force on a car than the car exerts on the truck—can interfere with learning. Furthermore, instructional design should be targeted at a level where students struggle appropriately and stay engaged in the learning process. The material should not be so unfamiliar and advanced that students become frustrated and disengage. The best PER-based curricula and pedagogies are designed to take into account the initial knowledge of a typical student and gradually build on it.

**Learning goals and assessment of learning**: Students learn what they are tested on. PER emphasizes the importance of having well-defined learning goals and assessing student learning using tools that are commensurate with those goals. "Students should understand acceleration" is not a well-defined goal because it does not make it clear to students what they should be able to do, whereas, "Students will compute the vector acceleration of an object from its definition, and use the result to reason conceptually about motion, and to solve quantitative problems using Newton's Second Law" gives a much clearer picture of what is expected. Examples of measurable goals should be shared with students and effective instruction should include students demonstrating how to apply physics concepts in diverse situations, analyzing problems by breaking them down into sub-problems, synthesizing a solution by combining different principles, and comparing and contrasting various concepts. PER-based approaches focus on helping all students develop a good conceptual understanding, and some approaches explicitly focus on integrating both conceptual and quantitative understanding. These instructional approaches bridge the gap between teaching and assessment of learning and provide feedback to both the instructor and students about what students have learned and their level of understanding throughout the course by using low-stakes built-in assessment tasks.[1]

**Learning is context-dependent:** Suppose your students learn about angular momentum conservation in the context of a spinning ballerina who speeds up when she brings her arms close to herself and slows down when she puts her arms out. They may not immediately discern the relevance of the same principle if the instructor assigns a problem in which a neutron star is collapsing under its own gravitational force and the student is asked to determine whether it will increase or decrease its rotational speed. To an instructor who has vast experience, recognizing that these two problems can be mapped onto each other is easy, but beginning students need effective instructional approaches to help them focus on the "deep" features of the problems.

**Student buy-in is important:** Since PER-based curricula and pedagogies keep students actively engaged in the learning process and make them struggle (appropriately) in order to develop expertise, obtaining buy-in from students is important. There should be an explicit discussion with the students of the fact that struggling is an integral part of learning, and the instructor is trying to help them get the most out of the course by using research-based approaches. Moreover, many introductory physics students believe that physics lacks a coherent knowledge structure and is a collection of facts and formulas or physics has nothing to do with their everyday life. Since such beliefs can impact motivation and learning, it is important to help students develop better attitudes and beliefs about learning physics.[5]

**Concluding remarks:** PER-based approaches are generally student-centered. They are designed to help all students but they are particularly helpful for underrepresented students and can decrease the gap between traditional and underrepresented students in physics classes.[6] Instructors have many PER-based approaches to choose from, and should find the ones that can be adopted or adapted to fit their particular instructional settings or constraints.[1] Before fully implementing any particular approach, it's best to try it first in some of the classes. Finally—don't forget to enjoy your teaching, inspire your students and help them see the big picture!